# A Comparative Study of Coherent and Incoherent Drives in Four-Level Quantum Dot Based Spaser

Ankit Purohit[1,*], and Akhilesh Kumar Mishra[2, *,†]
*Department of Physics, Indian Institute of Technology Roorkee, Roorkee-247667, Uttarakhand, India
†Centre for Photonics and Quantum Communication Technology, Indian Institute of Technology Roorkee, Roorkee- 247667, Uttarakhand, India
Email: [1]a_purohit@ph.iitr.ac.in, [2]akhilesh.mishra@ph.iitr.ac.in.

**Abstract**

In this article, we theoretically investigate a spaser (surface plasmon amplification by stimulated emission of radiation), which consists of a spherical silver nanoparticle surrounded by a four-level gain medium of quantum dots (QDs). The spaser system is pumped coherently and incoherently with the same excitation rate, and the characteristics of coherent localized surface plasmon (LSP) mode, thus produced, are compared for the two pumping scenarios. We provide a detailed analytical expression for the steady state and show that the incoherent pump is more suitable for the continuous spaser mode. The reason is better understood by studying the temporal evolution of the number of LSP ($N_n$), where the oscillation of LSP starts earlier for incoherent drive and relaxes to a steady state with a large value of $N_n$. At a large pump rate, the spaser curve shows saturation. In addition, we have found that the resonance peak of the spaser field is independent of coherent as well as incoherent pumping, while the peak amplitude of the field depends on the pump rate.

## 1. Introduction

Quantum optics and nano-plasmonics are rapidly expanding fields of study that offer new ways to generate and control the light at sub-wavelength dimension, utilize plasmons in quantum applications, and make miniaturized active plasmonic devices [1-5]. The confinement of light to a smaller volume can realize strong light-matter interaction. Localized surface plasmon (LSP) resonance is one such phenomenon wherein, under the presence of external time-dependent field, free electrons in a nano-size metal particle oscillate coherently. LSP is useful only when the resonance overcomes the metal's intrinsic losses. One of the ways to deal with the material losses is by transferring energy from external sources to sustain these resonances. The emission from these resonances can be intensified to behave like a laser since it exhibits characteristics analogous to those of a laser light [6-7]. A plasmonic nano-laser consists either of surface plasmon polariton mode or localized surface plasmon (LSP) mode [8]. But, a true spaser consists of LSP mode, which confines the field in all three-dimension, e.g., in a nanosized metal structure.

The idea of a spaser was proposed theoretically in 2003 by D. Bergman and M.I. Stockman [6]. Spaser is a nanoscale laser that typically comprises of subwavelength plasmonic cavity (metal nanoparticle (NP)) and a gain medium, which is usually semiconductor quantum dots (QDs). The gain medium can be excited optically, electrically, or chemically. The excited energy is directly transferred to the nearby LSP resonance, resulting in a macroscopic increase in the number of LSPs. Coherent LSP-based devices may have applications in nanoscale optoelectronic chips [3, 4]. The major drawbacks to realize the efficient spaser are high threshold and low number of LSP per spasing mode. To counter these, usually a large number of quantum dots had been used in experiments [9-11]. Although there is a theoretical study wherein a spaser using single quantum emitter has been realized but so far there has not been any experimental evidence [12].

The gain medium is usually considered as a two-level or three-level system in theory [7,13-15]. In a two-level gain medium, the gain saturation caused by feedback of SP mode limits the spaser operation [7]. K. E. Dorfman et al. showed that along with the incoherent pump, an extra coherent light source can increase the number of LSP and control the threshold in a three-level gain medium [13]. These are because of the quantum coherence and interference in three-level systems [4]. Theoretical and experimental developments show that quantum coherence-enhanced interaction gives way to several exotic phenomena such as electromagnetic-induced transparency, laser without inversion, ultraslow light and enhancement in refractive index [1].

The classical linear framework well describes the spaser threshold and spasing wavelength but is inadequate to describe the post threshold analysis due to gain saturation. Saturation introduces nonlinearity in Maxwell's equations written for a fixed frequency. The intensity-dependent dielectric function was later introduced to Maxwell's equations to deal with gain saturation [16, 17].

In this article, we use a quantum mechanical model to describe the spaser [7, 13-15]. We consider two scenarios to pump the gain medium. In one case, the pumping is incoherent, while in the other, it is considered coherent. The resulting dynamics have been studied and compared to get a deeper insight of the underlying phenomena. Usually, it is very cumbersome to achieve continuous spasing mode with coherent pumping in a two-level gain medium system because of the rapid population transfer in the energy levels of QDs. However, the use of coherent pumping makes it feasible to observe the spaser pulse in a three-level QD system [12]. So far, no study is available to compare the coherent and incoherent pumping effects on the continuous spaser mode. Since a four-level system is a typical system for dense QDs gain medium, we picked a four-level gain medium system for our investigation. In the following, we show the impact of coherent and incoherent pumps on the spasing curve, which is the variation of LSP population ($N_n$) with the input pump rate. For better understanding of the underlying dynamics, the time evolution in QDs and LSP are also studied.

## 2. Model and Mathematical framework

We consider a model in which a 40 nm silver nanoparticle is surrounded by a four-level QDs gain medium, as shown schematically in FIG. 1(a). The quantum dots (chromophore) energy levels $|0\rangle \rightarrow |3\rangle$ are coupled by external pump drive. For the first set of studies, we consider coherent pumping (external drive) of the QD levels with the rate $\Omega_g$ while in another set of studies we assume an incoherent pumping of the QD levels with a rate $g$, as depicted in FIG. 1(b). The population of level $|3\rangle$ decays to levels $|0\rangle$ and $|2\rangle$ with rates $\gamma_{30}$ and $\gamma_{32}$ respectively. Subsequently, the population from level $|2\rangle$ decays to level $|1\rangle$, which non-radiatively couples to the plasmon mode of the metal nanoparticle with Rabi frequency $\tilde{\Omega}_b$. The population in level $|1\rangle$ ultimately decays down to the ground level $|0\rangle$ to be again pumped to level $|3\rangle$. This operation with continuous input pumping ultimately reaches to the stationary spasing mode.

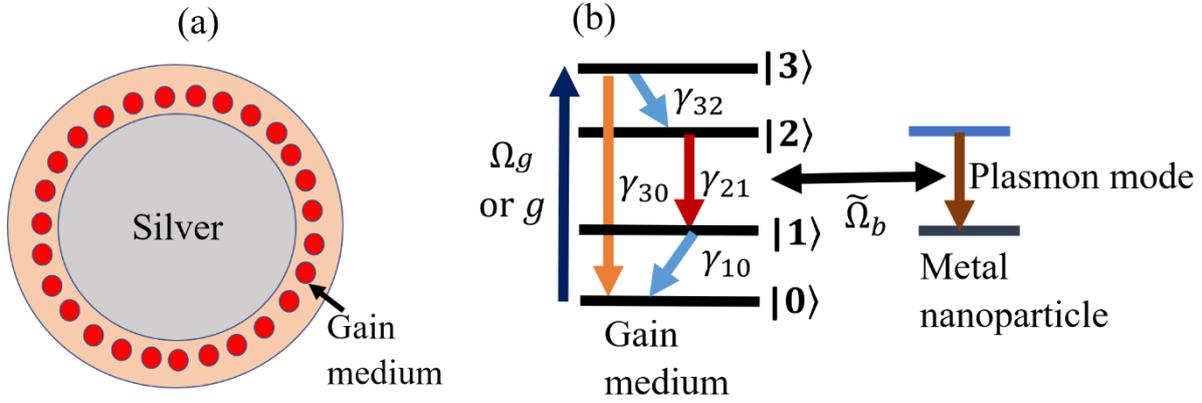

FIG. 1: (a) Schematic of spaser: A silver nanosphere is surrounded by homogeneous dielectric QDs gain medium. (b) Energy levels of four level gain medium where $|0\rangle$ and $|3\rangle$ levels are pumped coherently and incoherently with $\Omega_g$ and $g$ pump rates, respectively. Population transition between level $|2\rangle$ to level $|1\rangle$ stimulates the nearby LSP mode.

For the modelling, the quasi-classical model is adopted, in which gain medium is treated quantum mechanically, while the LSP and photons are treated semi-classically [7,13-15]. In our theory, we primarily use two approximations [18]. The first approximation ignores any correlation that may exist between the QDs. Such correlations are significant for small numbers of atoms or at extremely low temperatures. These correlations are critical for Bose-Einstein condensation and superconducting qubits. In the second approximation, all operators are assumed to behave as classical-numbers, which is valid when the number of atoms is large compared to one [1]. Therefore, the SP annihilation operator $\hat{a}_n$ can be expressed as a complex number and is written as $\hat{a}_n = a_{0n}e^{-i\omega t}$, where $a_{0n}$ is slowly varying amplitude of LSPs. The number of coherent LSPs per spasing mode is given by $N_n = |a_{0n}|^2$. The photon annihilation operator $\hat{b}_m$ is also a complex number, which is written as $\hat{b}_m = b_{0m}e^{-i\omega_m t}$, where $b_{0m}$ is slowly varying amplitude of photons.

The total Hamiltonian for the given system with rotating wave approximation is given as

$$H = \hbar\omega_n \hat{a}_n^\dagger \hat{a}_n + \sum_p \left[ \hbar\omega_m^p \hat{b}_m^\dagger \hat{b}_m + \sum_{i=0,1,2,3} \hbar\omega_i^p |i\rangle\langle i| + \hbar\left(\Omega_g^p |3\rangle\langle 0| + \Omega_b^p |2\rangle\langle 1| + c.c\right) \right], \quad (1)$$

where the first term is unperturbed Hamiltonian for the LSP mode, the second term represents unperturbed Hamiltonian for the pump source, third term is Hamiltonian for gain medium and last two terms are the interaction Hamiltonians that represent external coherent pump induced $|0\rangle \rightarrow |3\rangle$ transition with $\Omega_g$ strength and LSP mode interaction $|2\rangle \rightarrow |1\rangle$ transition with $\Omega_b$ strength, respectively. The Rabi frequencies for transition from $|2\rangle \rightarrow |1\rangle$ and $|0\rangle \rightarrow |3\rangle$ are given respectively by $\Omega_b = -A_n d_{21} \nabla\varphi_n(r_p) a_{0n}/\hbar = \tilde{\Omega}_b a_{0n}$, and $\Omega_g = E_m(r_p) d_{03} b_{0m}/\hbar$. In the analysis, we have considered $\Omega_g$ as constant i.e., independent of time because of continuous pumping for $|0\rangle \rightarrow |3\rangle$ transition.

The density matrix equation for single QD is given by Liouville-von Neumann equation, which is expressed as

$$\dot{\hat{\rho}} = -\frac{i}{\hbar}[H, \hat{\rho}] + L(\hat{\rho}), \qquad (2)$$

where the Lindblad term, which accounts for the dissipation and incoherent pump is written as

$$\begin{aligned} L(\hat{\rho}) = {} & \frac{\gamma_{32}}{2}\big(2\sigma_k\hat{\rho}\sigma_k^\dagger - \sigma_k^\dagger\sigma_k\hat{\rho} - \hat{\rho}\sigma_k^\dagger\sigma_k\big) + \frac{\gamma_{21}}{2}\big(2\sigma_n\hat{\rho}\sigma_n^\dagger - \sigma_n^\dagger\sigma_n\hat{\rho} - \hat{\rho}\sigma_n^\dagger\sigma_n\big) \\ & + \frac{\gamma_{10}}{2}\big(2\sigma_l\hat{\rho}\sigma_l^\dagger - \sigma_l^\dagger\sigma_l\hat{\rho} - \hat{\rho}\sigma_l^\dagger\sigma_l\big) + \frac{\gamma_{30}}{2}\big(2\sigma_m\hat{\rho}\sigma_m^\dagger - \sigma_m^\dagger\sigma_m\hat{\rho} - \hat{\rho}\sigma_m^\dagger\sigma_m\big) \\ & + \frac{g}{2}\big(2\sigma_m^\dagger\hat{\rho}\sigma_m - \sigma_m\sigma_m^\dagger\hat{\rho} - \hat{\rho}\sigma_m\sigma_m^\dagger\big), \end{aligned} \qquad (3)$$

where $\sigma_k = |2\rangle\langle 3|$, $\sigma_n = |1\rangle\langle 2|$, $\sigma_l = |0\rangle\langle 1|$, and $\sigma_m = |0\rangle\langle 3|$ are the transition operators for the four level QDs, $\gamma_{ij}$ is the spontaneous decay rate in respective $|i\rangle \to |j\rangle$ levels, and g is incoherent pump rate.

To make the Hamiltonian time independent, following transformation have been implemented $\hat{\rho}_{30} = \rho_{30}e^{-i\omega_{30}t}$, $\hat{b}_m = b_{0m}e^{-i\omega_{30}t}$, $\hat{\rho}_{21} = \rho_{21}e^{-i\omega t}$ and $\hat{a}_n = a_{0n}e^{-i\omega t}$, where $\rho_{30}$, $b_{0m}$, $\rho_{21}$, and $a_{0n}$ are slowly varying amplitudes.

The density matrix elements for any $p^{th}$ QD are written as:

$$\dot{\rho}_{11} = \gamma_{21}\rho_{22} - \gamma_{10}\rho_{11} - i[\Omega_b^*\rho_{21} - \Omega_b\rho_{21}^*], \qquad (4A)$$

$$\dot{\rho}_{22} = \gamma_{32}\rho_{33} - \gamma_{21}\rho_{22} - i[\Omega_b\rho_{21}^* - \Omega_b^*\rho_{21}], \qquad (4B)$$

$$\dot{\rho}_{33} = g\rho_{00} - (\gamma_{30} + \gamma_{32})\rho_{33} - i[\Omega_g\rho_{30}^* - \Omega_g^*\rho_{30}], \qquad (4C)$$

$$\dot{\rho}_{21} = i\Delta_b\rho_{21} - \frac{(\gamma_{21} + \gamma_{10})}{2}\rho_{21} - i\Omega_b[\rho_{11} - \rho_{22}], \qquad (4D)$$

$$\dot{\rho}_{30} = i\Delta_m\rho_{30} - \frac{(\gamma_{30} + \gamma_{32})}{2}\rho_{30} - i\Omega_g[\rho_{00} - \rho_{33}], \qquad (4E)$$

where $\Delta_m = \omega_m - \omega_{30}$ and $\Delta_b = \omega - \omega_{21}$ are detuning terms. The density matrix elements also satisfy the following conservation rule

$$\rho_{00} + \rho_{11} + \rho_{22} + \rho_{33} = 1. \qquad (4F)$$

The equation for LSP mode is written as:

$$\dot{a}_{0n} = -\Gamma_n a_{0n} - i\sum_p \rho_{21}\tilde{\Omega}_b, \qquad (5)$$

where $\Gamma_n = \gamma_n + i\Delta_n$, wherein $\gamma_n$ is SP relaxation rate, $\Delta_n = \omega_n - \omega$ is detuning term, and $\tilde{\Omega}_b = \Omega_b/a_{0n}$ is single plasmon Rabi frequency.

For the coherent pumping, we substitute the incoherent rate $g = 0$ in Eqs. (4A) - (4E). Whereas, for the incoherent pumping, we substitute $\Omega_g = 0$ in Eqs. (4A) - (4E). In addition, the coherent term $\rho_{30}$ or Eq. (4E) is not required for the analysis of the incoherent case.

## 3. Results and discussion

The general condition to establish spasing in the four-level system under consideration is $(\gamma_{32}, \gamma_{10}) \gg (\gamma_{21}, \tilde{\Omega}_b)$. From an experimental point of view, it is also critical that the spectrum of QDs must match the plasmon mode spectrum. For a 40 nm silver nanoparticle, we have considered the resonance frequency to be 2.50 eV, while the detuning parameter $\hbar(\omega_n - \omega)$ is 0.002 eV [13]. The decay rate for the plasmon mode is assumed to be $\gamma_n = 5.3 \times 10^{14} s^{-1}$. The following relaxation parameters, typical of semiconductor QDs, have been used in our numerical simulation: $\gamma_{21} = 4 \times 10^{12} s^{-1}$, $\gamma_{32} = 8 \times 10^{12} s^{-1}$, $\gamma_{10} = 8 \times 10^{12} s^{-1}$, $\gamma_{30} = 4 \times 10^{10} s^{-1}$ [19]. The number of quantum dots $\sum p$ i.e. $N_p$ is taken to be $6 \times 10^4$ [13]. The numerical value of the interaction strength between QDs and LSP mode is $\tilde{\Omega}_b^p = 4 \times 10^{12} s^{-1}$. Usually, the relaxation rates are not constant and vary with the higher order plasmon mode of metal sphere [20-22]. However, for the sake of simplification, we have considered all the decay rates to be constant [13,15]. We would like to note that the numerical values of detuning terms $\Delta_m = \omega_m - \omega_{30}$ and $\Delta_b = \omega - \omega_{21}$ are substituted to be zero.

### 3.1. Steady- state regime

FIG. 2 (b) displays the spaser curve, i.e., the variation of LSP population ($N_n$) with the input pump. The analytical and numerical simulation results are obtained by solving the Eqs. (4A) – (4F) and (5) in steady state ($\dot{\rho}_{ij} = 0$, $\dot{a}_{0n} = 0$) regime. Usually, the spaser curve is linear for two-level QDs, but that is not the case for three or four level QDs [7,13]. In a two level, once the threshold is reached, every quantum of excitation by pump would be stimulated and strongly correlated with the SP mode. In other words, there will be a linear dependence of $\rho_{22}$ or $N_n$ on the pump rate. However, in the presence of extra levels, this linear dependence is lost. Therefore, the resulting spaser curve is no longer linear. The solutions for $\rho_{22}$ for incoherent and coherent pump drives are given by Eq. (6) and (7), respectively, as expressed below

$$\rho_{22} = \left[1 + \left(\frac{\gamma_{10}(\gamma_{30} + \gamma_{32}) + g(\gamma_{10} + \gamma_{32})}{g\gamma_{32}}\right)\left(\frac{\Gamma_n(\gamma_{10} + \gamma_{21}) - i\Delta_b\Gamma_n}{2N_p\tilde{\Omega}_b^2}\right)\right]$$
$$\left[1 + \left(\frac{\gamma_{10}(\gamma_{30} + \gamma_{32}) + g(\gamma_{10} + \gamma_{32})}{g\gamma_{32}}\right)\right]^{-1}, \quad (6)$$

$$\rho_{22} = \left[1 + \left(\frac{\Gamma_n(\gamma_{10} + \gamma_{21}) - i\Delta_b\Gamma_n}{2N_p\tilde{\Omega}_b^2}\right)\left(1 + \frac{\gamma_{10}}{\gamma_{32}}\left(2 + \frac{(\gamma_{30} + \gamma_{32} - 2i\Delta_m)(\gamma_{30} + \gamma_{32})}{4\Omega_g^2}\right)\right)\right]$$
$$\left[1 + \left(1 + \frac{\gamma_{10}}{\gamma_{32}}\left(2 + \frac{(\gamma_{30} + \gamma_{32} - 2i\Delta_m)(\gamma_{30} + \gamma_{32})}{4\Omega_g^2}\right)\right)\right]^{-1}. \quad (7)$$

Eqs. (6) and (7) show that $\rho_{22}$ depends on the input pump rate, relaxation terms of QDs and LSP, number of QDs ($N_p$) and Rabi frequency $\tilde{\Omega}_b$. At the large input pump rate, $\rho_{22}$ saturates and saturation sets early for the coherent pump when compared with incoherent pump. $N_n$ follows similar saturation trend at large input pump as that of $\rho_{22}$.

$N_n$ depends linearly on the $\rho_{22}$ as given by the following equation,

$$N_n = \frac{1}{2\Gamma_n \tilde{\Omega}_b^2}\left[(\gamma_{21} - \gamma_{10})N_p \tilde{\Omega}_b^2 \rho_{22} + 0.5\Gamma_n \gamma_{10}(\gamma_{21} + \gamma_{10}) - i\Delta_b \Gamma_n\right]. \tag{8}$$

Hence, $N_n$ follows the same trends as $\rho_{22}$ with the input pump rate, as depicted in FIG. 2 (a) and (b). Note that we have considered only the spontaneous decay rate ($\gamma_{ij}$) and have not introduced the dephasing relaxation rate for transition starting from level $|2\rangle$. The incorporation of dephasing relaxation in level $|2\rangle$ will change the numerical values, but the trends with pump rate will not deviate from what is observed in FIG. 2.

FIG. 2(c) depicts the population inversion ($\rho_{22} - \rho_{11}$) variation with the pump rate. This figure shows that the spasing starts early for the incoherent pumping. It means that even for a very low input pump, a sufficient number of LSPs will be observed in steady state. Compared to the conventional laser, the inversion is much lower ($\sim 3.3 \times 10^{-3}$) for the spaser due to the strong feedback. On solving Eqs. (4D) and (5), we reach at Eq. (9), which reveals that the inversion is independent of the input pump rate once the steady state sets in, as is also shown in FIG. 2(c).

$$\rho_{22} - \rho_{11} = \frac{\Gamma_n(\gamma_{21} + \gamma_{10}) - i\Delta_b \Gamma_n}{2N_p \tilde{\Omega}_b^2} \tag{9}$$

Inset of FIG. 2(c) shows the variation of $\rho_{22} - \rho_{11}$ with very high input pump rate. The population inversion remains independent of the pump rate once the steady state is reached.

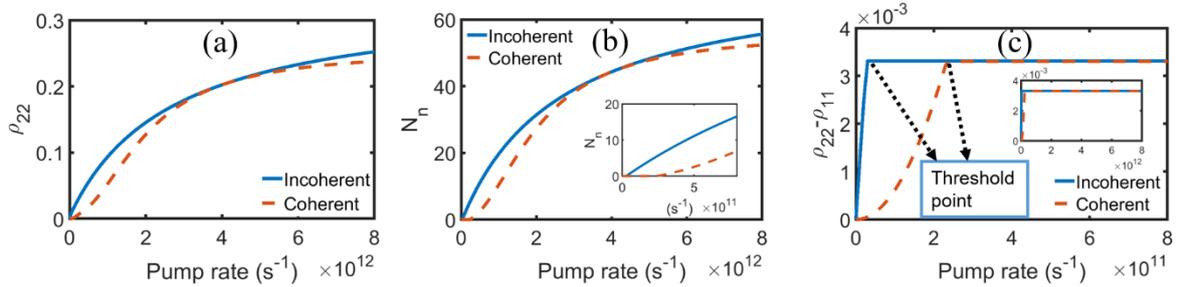

FIG. 2: Steady-state properties of spaser curve for coherent and incoherent pump drive. (a) Variation of $\rho_{22}$ with the pump rate, (b) Variation of number of LSP, $N_n$ as function of pump rate. Inset in FIG. 2(b) shows the evolution of $N_n$ at small pump rate. (c) Variation of population inversion, $\rho_{22} - \rho_{11}$, as a function of pump rate. Inset in FIG.2 (c) show the evolution of $\rho_{22} - \rho_{11}$ for larger pump rate.

The spasing frequency can be calculated by equating the imaginary part of Eq. (9) to zero, which gives

$$\omega = \frac{\omega_n(\gamma_{21} + \gamma_{10}) + \omega_{21}\gamma_n}{\gamma_{21} + \gamma_{10} + \gamma_n}. \tag{10}$$

For two level system, the spasing frequency depends only on $\gamma_{21}$ and $\gamma_n$, but for four level system it depends on $\gamma_{10}$ as well [7].

### 3.2 Transient regime

To gain a better insight into the continuous spaser mode, FIG. 3 (a) and (b) display the time evolution of $N_n$ at the input pump rates of $2 \times 10^{12} s^{-1}$ and $4 \times 10^{12} s^{-1}$, respectively. Eqs.

(5) and (4D) show that the temporal profile of $N_n$ is related to the temporal profiles of $\rho_{22}$ and $\rho_{11}$. Therefore, in FIG. 4, we have plotted the temporal profiles of $\rho_{00}$, $\rho_{11}$, $\rho_{22}$, and $\rho_{33}$ at the input pump rate $2 \times 10^{12} s^{-1}$. These numerical simulations show that $N_n$ starts oscillation at same time as $\rho_{22}$ does and they both relax to the steady state almost simultaneously. At small pump rate of $2 \times 10^{12} s^{-1}$, $N_n$ initially shows high amplitude oscillations that relaxes to steady-state after ~2.5 ps. As the pump rate increases to $4 \times 10^{12} s^{-1}$, the number of oscillations, as well as the amplitude, increases further, which takes longer to relax to steady-state and contributes a relatively larger number of LSPs in spasing mode. Since the oscillations start early for incoherent pumping, incoherent pump is preferred in conventional lasers. However, the coherent pumping is useful in certain cases such as in spaser pulse [12,14]. Hence, the onset of oscillations at very small pump rate is the reason of early threshold for incoherent pump drive as shown in FIG. 2(c). The insets of FIG. 3(a) and 3(b) shows the steady state values of $N_n$ at the pump rates of $2 \times 10^{12} s^{-1}$ and $4 \times 10^{12} s^{-1}$, respectively. At $2 \times 10^{12} s^{-1}$, different values of $N_n$ are observed for coherent and incoherent pumping, while almost same value of $N_n$ is observed at the $4 \times 10^{12} s^{-1}$. Same can be observed for Fig. 2(b).

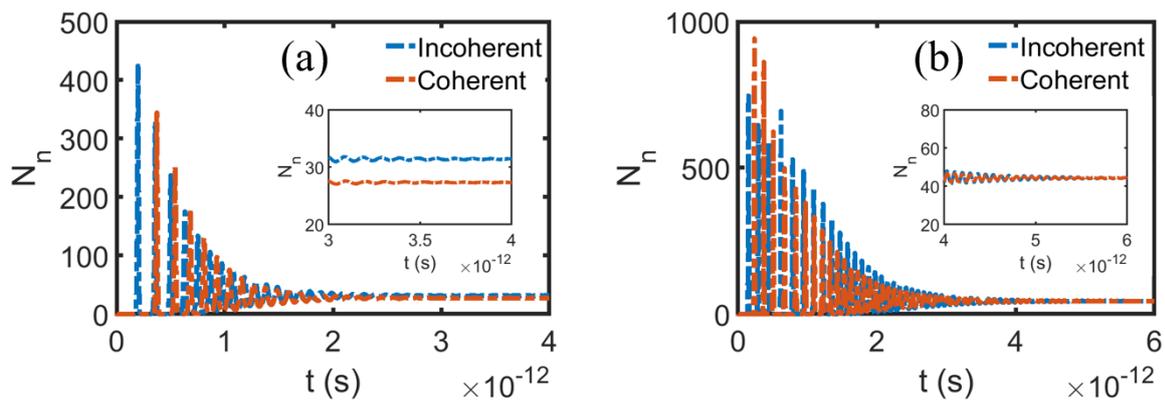

FIG. 3: Temporal profile of $N_n$ for coherent and incoherent pump at (a) $2 \times 10^{12} s^{-1}$ and (b) $4 \times 10^{12} s^{-1}$ rate. Inset shows the steady-state population of LSP ($N_n$) for respective pump rates. Insets in FIGS. 3(a) and 3(b) show steady state values of $N_n$ for $2 \times 10^{12} s^{-1}$ and $4 \times 10^{12} s^{-1}$ rates respectively.

FIG. 4 (a) depicts that $\rho_{00}$ is relatively small with an incoherent pump of rate of $2 \times 10^{12} s^{-1}$. However, the populations in the upper levels ($\rho_{11}, \rho_{22}, and\ \rho_{33}$) are larger with the incoherent pump. Population oscillations start in the active level ($\rho_{11}\ and\ \rho_{22}$) due to coupling with SP mode, and with time, they relax down to steady state. As mentioned above, these oscillations are the reason behind the oscillation in $N_n$ as depicted in FIG. 3. At a higher pumping rate ($4 \times 10^{12} s^{-1}$), similar oscillations appear in the active levels, but they relax to the same steady value irrespective of the incoherent and coherent drives (figure not shown). As a result, we witness the same number of LSP for spasing mode at the rate of $4 \times 10^{12} s^{-1}$, as shown in FIG. 2(b) and FIG. 3(b). As shown in FIG. 4 (d), the temporal evolution depicts that $\rho_{33}$ decays quickly with the coherent pumping due to the extra coherence ($\rho_{30}$) term in Eq. (4E).

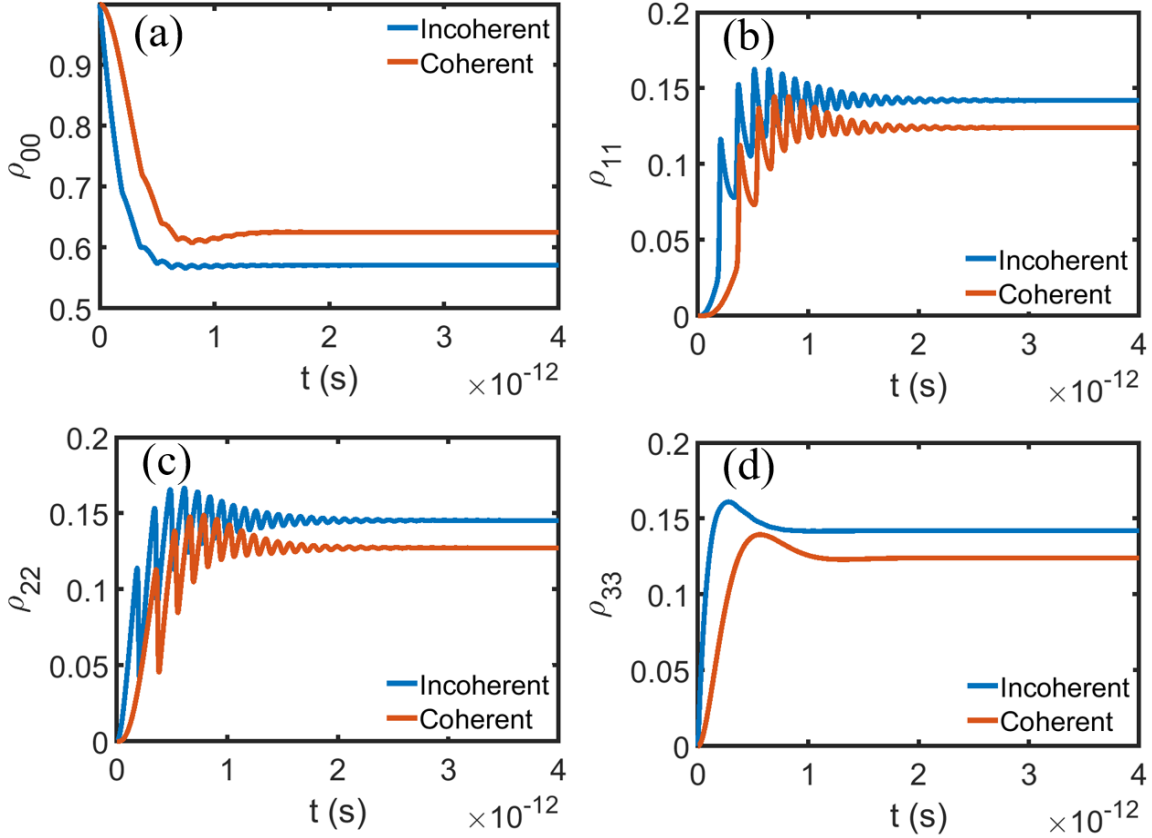

FIG. 4: Temporal profile of different level populations. Temporal dynamics of (a) $\rho_{00}$, (b) $\rho_{11}$, (c) $\rho_{22}$, (d) $\rho_{33}$ at pump rate $2 \times 10^{12} s^{-1}$.

The spaser field, as expressed by Eqn. (11), is determined by taking the Fourier transform of $a_{0n}(t)$ −

$$E(\omega) = \int_0^T a_{0n}(t) \, exp(i\omega_0 t) dt \qquad (11)$$

FIG. 5 (a) and (b) display the spaser field amplitude $|E(\omega)|$ for the pump rates of $2 \times 10^{12} \, s^{-1}$ and $4 \times 10^{12} \, s^{-1}$, respectively. We choose $T \sim 4 \times 10^{12} \, s$, so that it can include all the oscillations of FIG. 3(a) and 3(b). As shown in FIG. 5(a), the peak amplitude of the spaser field for incoherent pump is larger compared to that for coherent pump. Although at pump rate $4 \times 10^{12} \, s^{-1}$, same peak amplitude is achieved for both coherent and incoherent pumps. The spasing frequency is given in Eq. (10), which depends on the metal nanoparticle, gain medium resonance frequency and decay rate but does not depend on the pump rate. In addition to the resonance peak, the spaser field demonstrates many non-resonant frequency oscillations, similar to the conventional laser. As per the laser or spaser theory, the non-resonance frequency terms mainly depend on the spontaneous emission rate (or the decay rates) of the gain medium. As we can see in Eq. (4C) and (4E), an extra coherence term $\rho_{30}$ is present for the coherent pump drive, which increases the decay rates of the gain medium compared to that for the incoherent pump drive case. Therefore, for a particular pump rate, the non-resonant frequency components for the coherent and incoherent drives are slightly different.

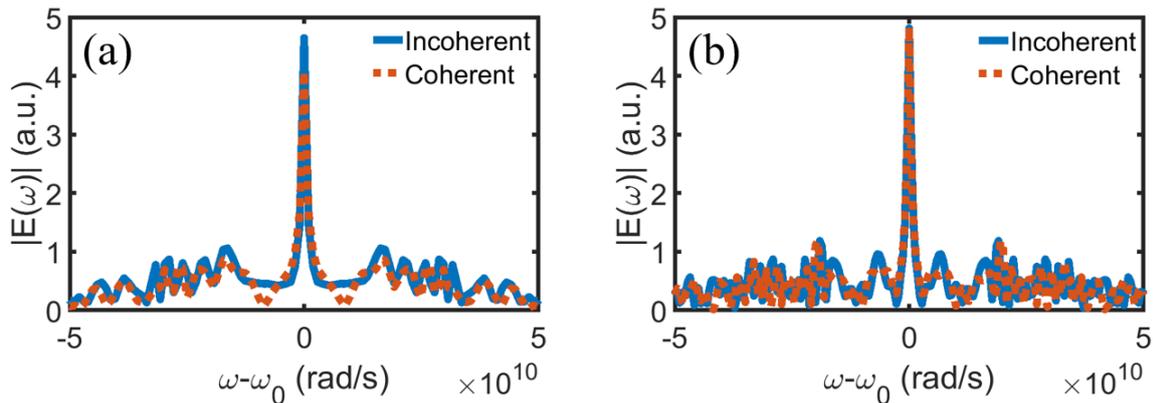

FIG. 5: Spaser field for coherent and incoherent pump drives at (a) $2 \times 10^{12} s^{-1}$ (b) $4 \times 10^{12} s^{-1}$ rate.

The quality factor of a plasmonic cavity is given by [23]

$$Q = \frac{\omega}{\omega_s} = \frac{v}{\gamma_s},  \quad (12)$$

where $v$ is spasing frequency and $\gamma_s$ is the spaser linewidth.

The quality factor for a cavity is also defined as [23]

$$Q = 2\pi v \frac{\text{energy stored in a cavity}}{\text{energy lost per unit time}} = 2\pi v \frac{P_{out} t_c}{(hv/t_c)}, \quad (13)$$

where $P_{out} (= hv\gamma_n N_n)$ is absorption power for the metal nanoparticle and $t_c = 1/(2\pi\gamma_{21})$ is cavity lifetime [22, 23]. Note that we consider only the fundamental linewidth of spaser which arises due to the spontaneous emission rate $(\gamma_{21})$ of the gain medium.

The substitution of (12) to (13) gives spaser linewidth [23],

$$\gamma_s = \frac{2\pi\gamma_{21}^2}{\gamma_n N_n}. \quad (14)$$

which depends on the gain medium decay rate $\gamma_{21}$, LSP decay rate $\gamma_n$ and number of LSP $N_n$. Parameters $\gamma_{21}$ and $\gamma_n$ are constant for the coherent and incoherent pump drives and $N_n$ is the only variable. For the LSP, the difference in magnitudes of $N_n$ for the coherent and incoherent drives is very small for a given pump rate. Hence, the spectral width for the spaser, which largely depends on the decay rates of LSP and the gain medium, remains almost the same for coherent and incoherent drives.

The theoretical expression of linewidth for coherent and incoherent pumping can be derived by substituting the values of $N_n$ from equations (6), (7), and (8) into the equation (14). Fig. 6 illustrates the variation of linewidth with the input pump rate. This figure indicates that the spaser linewidth narrows as the pump rate increases [22]. Near the threshold pump rate, a significant difference in spaser linewidths is observed for coherent and incoherent pumping, whereas at high pump rates, the difference is diminished.

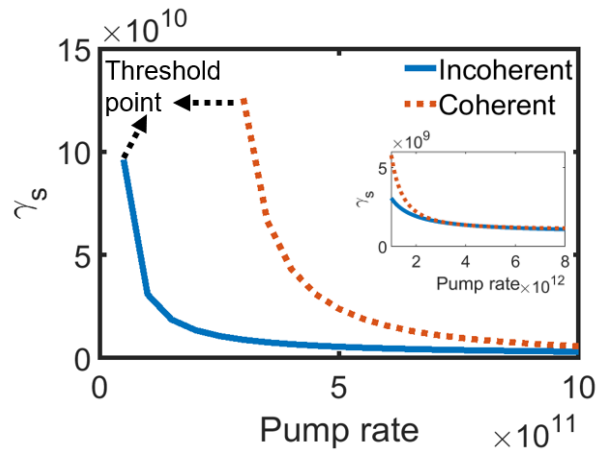

FIG. 6. Spaser linewidth with low pump rate. Inset shows the spaser linewidth at high pump rate.

It is important to note that this study focuses on the LSP that is characterized by very high decay rate ($\gamma_n \sim 10^{14} s^{-1}$). We would like to emphasize that for an SPP laser or conventional laser, which has lower dissipation rate and large number of SPP or photon, coherent and incoherent pumping will have appreciable difference.

## 4. Conclusion

In summary, we studied the continuous spaser mode system with a four-level QDs gain medium pumped coherently and incoherently. With coherent pumping, the coherence established between levels $|0\rangle$ and $|3\rangle$ contributes negatively to the number of LSP, $N_n$. However, at a particular pump rate $\sim 4 \times 10^{12} s^{-1}$, we observed same number of LSP for both incoherent as well as coherent pump drives. $N_n$ follows the same trend as $\rho_{22}$ with continuous input pump rate. The threshold rate is reached early for the incoherent pump and we observed a larger number of LSPs. These results explain the preference of incoherent pumping over coherent pumping in conventional laser for the dense QDs gain medium. The temporal profiles of $\rho_{22}$ and $N_n$ provide insight into the steady-state results. It has been shown that generally the oscillations in occupational probabilities of different levels start early and last longer for incoherent pump. As a result, a larger number of LSPs are produced with incoherent pumping. The initial oscillations in $N_n$, $\rho_{11}$ and $\rho_{22}$ are due to the mutual perturbation between $N_n$ and the gain medium's $|2\rangle \rightarrow |1\rangle$ transition. The numerical plot of spaser field shows that the peak amplitude is slightly larger for the incoherent pump rate of $2 \times 10^{12} s^{-1}$, while the peak amplitudes are almost same for $4 \times 10^{12} s^{-1}$ pump rate. The spaser field is accompanied by a larger number of non-resonant frequency oscillations, which are typical of any laser system. We believe these results provide a better understanding of the spaser system.

**Data availability statement-**The data that support the findings of this study are available from the corresponding author upon reasonable request.

**Acknowledgement**

Ankit Purohit wishes to acknowledge university grants commission (UGC), India for the research fellowship.

**References**

[1] M. O. Scully, and M. S. Zubairy, "Quantum optics", Cambridge university press, *Cambridge, CB2 2RU, UK* (1997).
[2] Howard Carmichael, "An open quantum system approach to quantum optics", *Springer-Verlag,* (1991).
[3] Palash Bharadwaj, Bradley Deutsch, and Lukas Novotny. "Optical antennas." *Advances in Optics and Photonics* 1, no. 3 (2009): 438-483.
[4] Shangjr Gwo, and Chih-Kang Shih. "Semiconductor plasmonic nanolasers: current status and perspectives." *Reports on Progress in Physics* 79, no. 8 (2016): 086501.
[5] D. S. Dovzhenko, S. V. Ryabchuk, Yu P. Rakovich, and I. R. Nabiev. "Light–matter interaction in the strong coupling regime: configurations, conditions, and applications." *Nanoscale* 10, no. 8 (2018): 3589-3605.
[6] David J. Bergman, and Mark I. Stockman. "Surface plasmon amplification by stimulated emission of radiation: quantum generation of coherent surface plasmons in nanosystems." *Physical review letters* 90, no. 2 (2003): 027402.
[7] Mark I. Stockman, "The spaser as a nanoscale quantum generator and ultrafast amplifier." *Journal of Optics* 12, no. 2 (2010): 024004.
[8] Shaimaa I. Azzam, Alexander V. Kildishev, Ren-Min Ma, Cun-Zheng Ning, Rupert Oulton, Vladimir M. Shalaev, Mark I. Stockman, Jia-Lu Xu, and Xiang Zhang. "Ten years of spasers and plasmonic nanolasers." *Light: Science & Applications* 9, no. 1 (2020): 90.
[9] M. A. Noginov, G. Zhu, A. M. Belgrave, Reuben Bakker, V. M. Shalaev, E. E. Narimanov, S. Stout, E. Herz, T. Suteewong, and U. Wiesner. "Demonstration of a spaser-based nanolaser." *Nature* 460, no. 7259 (2009): 1110-1112.
[10] Pei Song, Jian-Hua Wang, Miao Zhang, Fan Yang, Hai-Jie Lu, Bin Kang, Jing-Juan Xu, and Hong-Yuan Chen. "Three-level spaser for next-generation luminescent nanoprobe." *Science advances* 4, no. 8 (2018): eaat0292.
[11] Ekaterina I. Galanzha, Robert Weingold, Dmitry A. Nedosekin, Mustafa Sarimollaoglu, Jacqueline Nolan, Walter Harrington, Alexander S. Kuchyanov et al. "Spaser as a biological probe." *Nature communications* 8, no. 1 (2017): 15528.
[12] Pu Zhang, Igor Protsenko, Vahid Sandoghdar, and Xue-Wen Chen. "A single-emitter gain medium for bright coherent radiation from a plasmonic nanoresonator." *Acs Photonics* 4, no. 11 (2017): 2738-2744.
[13] Konstantin E. Dorfman, Pankaj K. Jha, Dmitri V. Voronine, Patrice Genevet, Federico Capasso, and Marlan O. Scully. "Quantum-coherence-enhanced surface plasmon amplification by stimulated emission of radiation." *Physical review letters* 111, no. 4 (2013): 043601.
[14] Pankaj K. Jha, Yuan Wang, Xuexin Ren, and Xiang Zhang. "Quantum-coherence-enhanced transient surface plasmon lasing." *Journal of Optics* 19, no. 5 (2017): 054002.
[15] Lakshitha Kumarapperuma, Malin Premaratne, Pankaj K. Jha, Mark I. Stockman, and Govind P. Agrawal. "Complete characterization of the spasing (ll) curve of a three-level quantum coherence enhanced spaser for design optimization." *Applied Physics Letters* 112, no. 20 (2018).
[16] Si-Yun Liu, Jiafang Li, Fei Zhou, Lin Gan, and Zhi-Yuan Li. "Efficient surface plasmon amplification from gain-assisted gold nanorods." *Optics letters* 36, no. 7 (2011): 1296-1298.

[17] Nikita Arnold, Klaus Piglmayer, Alexander V. Kildishev, and Thomas A. Klar. "Spasers with retardation and gain saturation: electrodynamic description of fields and optical cross-sections." *Optical Materials Express* 5, no. 11 (2015): 2546-2577.

[18] A. A. Zyablovsky, E. S. Andrianov, I. A. Nechepurenko, A. V. Dorofeenko, A. A. Pukhov, and A. P. Vinogradov. "Approach for describing spatial dynamics of quantum light-matter interaction in dispersive dissipative media." *Physical Review A* 95, no. 5 (2017): 053835.

[19] Christos Tserkezis, Christian Wolff, Fedor A. Shuklin, Francesco Todisco, Mikkel H. Eriksen, P. A. D. Gonçalves, and N. Asger Mortensen. "Gain-compensated cavities for the dynamic control of light-matter interactions." *Physical Review A* 107, no. 4 (2023): 043707.

[20] Günter Kewes, Kathrin Herrmann, Rogelio Rodríguez-Oliveros, Alexander Kuhlicke, Oliver Benson, and Kurt Busch. "Limitations of particle-based spasers." *Physical Review Letters* 118, no. 23 (2017): 237402.

[21] E. S. Andrianov, A. A. Pukhov, A. V. Dorofeenko, A. P. Vinogradov, and A. A. Lisyansky. "Rabi oscillations in spasers during nonradiative plasmon excitation." *Physical Review B* 85, no. 3 (2012): 035405.

[22] Mark I. Stockman "Brief history of spaser from conception to the future." *Advanced Photonics* 2, no. 5 (2020): 054002-054002.

[23] A. Ghatak, and K. Thyagarajan, "Optical electronics" Cambridge university press, (1989).